# Analytical modelling of thermal residual stresses and optimal design of $ZrO_2$/($ZrO_2$+Ni) sandwich ceramics

Wenbin Zhou[a,*] Rubing Zhang[b,*], Shigang Ai[b], Yongmao Pei[a], and Daining Fang[a]

[a]*State Key Lab for Turbulence and Complex Systems, Peking University, Beijing, 100871, China*

[b]*Department of Mechanics, Beijing Jiaotong University, Beijing, 100044, China*

**Abstract**

The joining of ceramics with metals have been extensively used in applications requiring high strength and excellent heat insulation. However, evaluating the residual stress generated inevitably due to the mismatch in coefficients of thermal expansion of ceramic and metal is challenging, which is very important for fabrication and characterization of layered inhomogeneous material. A simplified analytical model considering the overall deformation compatibility is established to compute the interlaminar residual stresses of the $ZrO_2$/($ZrO_2$+Ni) sandwich ceramics, which agrees well with the results obtained by the commercial finite element package. The effects of the thickness ratio of the transitional layer to the middle layer, and the number of transitional layers on the properties of the $ZrO_2$/($ZrO_2$+Ni) sandwich ceramics are researched to obtain the optimal structure.

## 1. Introduction

Zirconia ceramic, as one of the most widely used structural ceramics, has attracted considerable attention because of its excellent combination of physicochemical properties, such as high melting point and toughness due to stress-induced phase transformation, low thermal conductivity and outstanding physical and chemical stability at high temperature. These excellent properties make $ZrO_2$ ceramic a potential candidate for a variety of structural and multifunctional applications, including solid oxide fuel cells, oxygen sensors, ceramic membranes, and light-weight parts used at high temperature. However, in spite of lots of excellent properties for different applications, the brittle and defect sensitive instinct of structural ceramics has long prevented $ZrO_2$ ceramic from being more widely used. Many attempts have been made to solve such problems, mainly by introducing a toughening phase such as particles, metal, fibers and whiskers. Unfortunately, the unsatisfactory fracture toughness improvement by above methods is still the obstacle to a wider range of use, especially for applications in severe environments. Recently, multilayer structure design such as ceramic-metal sandwich structure is considered as one of the effective strategies to increase toughness and improve ceramic performance. For example, EL-Wazery et al. [1] experimentally investigated the microstructure and mechanical properties of TZ3Y/Ni ceramic-metal materials fabricated by powder metallurgy technique and found that the fracture toughness and the elastic modulus were greatly improved. However, this introduces interfaces and results in thermal mismatch. If the property mismatch across interfaces is large, then this leads to large thermal stresses that are undesirable. Therefore, it is of great importance to understand and be able to predict the distribution and magnitude of thermal residual stresses in the ceramic-metal jointing materials.

---

* Corresponding author. Tel.: +86 010 62757417
Email address: wbzhou@pku.edu.cn and rbzhang@bjtu.edu.cn

Studies on the effects of thermal residual stresses in ceramic-metal jointing systems with various geometries have been performed extensively. Residual stresses in multi-layer and compositionally graded plates have been studied by Ravichandran [2] for $Ni-Al_2O_3$ systems, by Zhao et al. [3] for $Al_2O_3/(W,Ti)C$ systems, and by Lannutti et al. [4] for $NiAl-Al_2O_3$ systems. In addition to plate geometries, coating and joint geometries have also been investigated. These include the work conducted by Baig et al. [5] for $ZrO_2$ coatings on Nimonic substrates, by Drake et al. [6] and He et al. [7] for joint geometries, and by Itoh and Kashiwaya [8] for both coatings and joints. Based on the basic assumptions of the shear-lag theory, Nairn and Mendels [9] proposed an optimal shear-lag method for the plane stress problems of the layered material, which can give the analytical expression of the interlaminar shear stress and the residual normal stress of each layer of the multi-layer materials. Suhir [10] established the theoretical model of residual stresses in multi-layer material interface with the theory of bending stress, and the distribution formulae of stress such as shear stress, normal stress and peel stress were deduced. These models have great significance in predicting residual stresses theoretically, however, they both have some ideal assumptions, which greatly limited their ranges of application. For example, Nairn assumed that the peel stress perpendicular to the interface is zero, and Suhir assumed that at least one of the multi-layer components is thick and stiff enough, so that this component and the assembly as a whole do not experience bending deformations.

In this study, a simplified analytical model considering the overall deformation compatibility has been established to compute the interlaminar residual stresses of the porous $ZrO_2$/($ZrO_2$+Ni) sandwich ceramics, which are fabricated by cold isostatic pressing and pressureless sintering (CIP-PLS). Microstructures and properties of the porous $ZrO_2$/($ZrO_2$+Ni) sandwich ceramics were experimentally studied in our previous work[11-12]. Then a finite element model is established to verify the correctness of the analytical model. The residual stress field characterized by the analytical model quantitatively agrees well with the finite element method. At last, an optimal design has been done on the thermal bridge block structure, the effects of the thickness ratio of the transitional layer to the middle layer, and the number of transitional layers on the mechanical properties are researched.

## 2. Residual stress analysis model

### 2.1. Numerical simulation model

As we can see from the above analysis, the residual stresses produced during the fabrication process have a great effect on the mechanical properties of $ZrO_2$/($ZrO_2$+Ni) sandwich ceramics, especially the interfacial stresses. Therefore, it is of significance to study the interfacial stresses. Fig.1 shows the schematic composition and optical photograph of the $ZrO_2$/($ZrO_2$+Ni) sandwich ceramics fabricated by cold isostatic pressing and pressureless sintering (CIP-PLS). The samples of 34 mm×5 mm×6 mm are cut off from the as-sintered ceramics in this work, with $h_1$=9mm, $h_2$=3mm, and $h_3$=10mm in the thickness direction. Fig.2 shows the finite element model of the $ZrO_2$/($ZrO_2$+Ni) sandwich ceramics obtained by the commercial finite element package ABAQUS. The specimens are assumed to be stress free at the temperature of 1200 ℃ (i.e., reference temperature), at which the spraying process is assumed to end [13]. To all analyzed systems, the final residual stresses are only generated due to the cooling of the whole coating specimens from the reference temperature to room temperature (25 ℃). Here, the FE model is the same size as the fabricated specimen. The mesh division is performed with 78,040 8-node hex elements (C3D8R). The material properties are shown in Table1[14-16], where the material constants in each composition for finite element analysis, such as Young's modulus, coefficient of thermal expansion (CTE) and Poisson's ratio, are calculated from the mixing rule below [17-18]:

$$\left. \begin{aligned} E &= \frac{8E_0(1-P)(1+\mu_0)(23+8\mu_0)}{8(1+\mu_0)(23+8\mu_0)+P(5+\mu_0)(37-8\mu_0)} \\ \alpha &= \alpha_0 \quad , \quad \mu = \mu_0 \end{aligned} \right\} \tag{1}$$

where $P$ is the porosity, $E_0$, $\alpha_0$ and $\mu_0$ are all dependent on the ceramic volume fraction and can be expressed from the Vegard's rule below [19]

$$M_0 = M_c V_c + M_m V_m \qquad (2)$$

where $M$ represents the material property parameters ($E$, $\alpha$ and $\mu$), $V_c$ donates the ceramic volume fraction, and subscript $c$ and $m$ represent ceramic and metal, respectively.

## 2.2. Theoretical model

The simplified stress analysis model is shown in Fig.3a. Firstly we only consider a two-layer structure which was fabricated at an elevated temperature and subsequently cooled. The longitudinal displacements $u_1(y)$ of layer 1 and $u_2(y)$ of layer 2 can be expressed as follows [10]:

$$\left.\begin{array}{l} u_1(y) = -\alpha_1 \Delta t y + \lambda_1 \int_0^y T(\xi) d\xi - \kappa_1 \tau(y) \\ u_2(y) = -\alpha_2 \Delta t y - \lambda_2 \int_0^y T(\xi) d\xi + \kappa_2 \tau(y) \end{array}\right\} \qquad (3)$$

where $\tau(y)$ is the shear stress in the interface,

$$T(y) = \int_{-l}^y \tau(\xi) d\xi \qquad (4)$$

is the shear force per unit width for the given cross section y. $l$ is half the length. $\alpha_1, \alpha_2$ are thermal expansion coefficients of the layer 1 and layer 2, respectively. $\lambda_i = \dfrac{1-\nu_i}{E_i h_i}$ is coefficients of axial compliance for layer i, $\kappa_i = \dfrac{2(1+\nu_i)h_i}{3E_i}$ is coefficients of interfacial compliance [10], $E_i$ is elastic modulus, $\nu_i$ is poisson's ratio, $\Delta t$ is the temperature differential.

Using the condition $u_1(y) = u_2(y)$ of the displacement compatibility, we obtain the following equation for the unknown shear stress $\tau(y)$:

$$\tau(y) - k^2 \int_0^y T(\xi) d\xi = \frac{\Delta \alpha \Delta t}{k} y \qquad (5)$$

where $\kappa = \sum \kappa_i$, $\lambda = \sum \lambda_i$, $k = \sqrt{\dfrac{\lambda}{\kappa}}$.

The equation (5) has the following solution:

$$\tau(y) = \frac{k}{\lambda} \Delta \alpha \Delta t \frac{sinh(ky)}{cosh(kl)} = \frac{\Delta \alpha \Delta t}{\lambda} \chi_1(y) \qquad (6)$$

where $\chi_1(y) = k \dfrac{sinh(ky)}{cosh(kl)}$ characterizes the distribution of the shearing stress.

We can see that in the above two-layer model, the shear stress is only related to the adjacent two layers, and when the layer is more than two layers, the model becomes very complicated.

Here we propose a simple model considering the whole structure's displacement compatibility to compute the interfacial stresses in multi-layer structure. Based on the above model and our FE results, we assume the shear stresses have the following distribution characteristic:

$$\tau_1(y) = \tau_{10} \chi_1(y) \ , \ \tau_2(y) = \tau_{20} \chi_1(y) \qquad (7)$$

where $\chi_1(y) = k \dfrac{sinh(ky)}{cosh(kl)}$ characterizes the distribution of the shearing stresses, $\tau_{10}$ and $\tau_{20}$ characterize the magnitudes of the shear stresses, which are decided by the whole structure's displacement compatibility condition. The shear forces per unit width for the given cross section $y$ are

$$T_1(y) = \int_{-l}^{y} \tau_1(\xi) d\xi = -\tau_{10} \chi_0(y) \ , \ T_2(y) = \int_{-l}^{y} \tau_2(\xi) d\xi = -\tau_{20} \chi_0(y) \qquad (8)$$

where $\chi_0(y) = 1 - \dfrac{cosh(ky)}{cosh(kl)}$ characterizes the distribution of the force $T(y)$. Note that $\chi_1(y) = -(d\chi_0(y))/(dy)$.

The equilibrium equations for the force are given by

$$\left. \begin{array}{l} \sigma_1(y) = -T_1(y)/h_1 = \dfrac{\tau_{10}}{h_1} \chi_0(y) \\[2mm] \sigma_2(y) = (T_1(y) - T_2(y))/h_2 = \dfrac{-(\tau_{10} - \tau_{20})}{h_2} \chi_0(y) \\[2mm] \sigma_3(y) = T_2(y)/h_3 = \dfrac{-\tau_{20}}{h_3} \chi_0(y) \end{array} \right\} \qquad (9)$$

$$\int_{-l}^{l} q_1(\xi) d\xi = 0 \ , \ \int_{-l}^{l} q_2(\xi) d\xi = 0 \qquad (10)$$

The equilibrium equations for the portion are expressed as

$$\left. \begin{array}{l} \displaystyle\int_{-l}^{y}\int_{-l}^{\xi} q_1(\xi) d\xi d\xi = T_1(y)\dfrac{h_1}{2} \\[4mm] \displaystyle\int_{-l}^{y}\int_{-l}^{\xi} q_2(\xi) d\xi d\xi = -T_1(y)\dfrac{h_1}{2} + T_1(y)\dfrac{h_2}{2} + T_2(y)\dfrac{h_2}{2} \end{array} \right\} \qquad (11)$$

Combing Eqs. (10) and (11), we get the following interface peel stresses:

$$\left.\begin{array}{l} q_1(y) = \dfrac{k^2 h_1 \tau_{10}}{2} \dfrac{cosh(ky)}{cosh(kl)} - \dfrac{k h_1 \tau_{10}}{2l} \tan h(kl) \\[4mm] q_2(y) = \dfrac{k^2(-h_1 \tau_{10} + h_2 \tau_{10} + h_2 \tau_{20})}{2} \dfrac{cosh(ky)}{cosh(kl)} - \dfrac{k(-h_1 \tau_{10} + h_2 \tau_{10} + h_2 \tau_{20})}{2l} \tan h(kl) \end{array}\right\} \tag{12}$$

The whole structure's displacement compatibility condition (see Fig.3b) is

$$\left.\begin{array}{l} \Delta_1 + \Delta_2 = (\alpha_1 - \alpha_2)\Delta t l \\[2mm] \Delta_1 + \Delta_3 = (\alpha_1 - \alpha_3)\Delta t l \end{array}\right\} \tag{13}$$

where $\Delta_1, \Delta_2, \Delta_3$ are the deformation of each layer under the residual stresses and can be calculated as follows:

$$\left.\begin{array}{l} \Delta_1 = 2\displaystyle\int_0^l \dfrac{\sigma_1(y)}{E_1} dy = \dfrac{2(l - tanhkl)}{E_1 h_1} \tau_{10} = A_1 \tau_{10} \\[4mm] \Delta_2 = -2\displaystyle\int_0^l \dfrac{\sigma_2(y)}{E_2} dy = \dfrac{2(l - tanhkl)}{E_2 h_2}(\tau_{10} - \tau_{20}) = A_2(\tau_{10} - \tau_{20}) \\[4mm] \Delta_3 = -2\displaystyle\int_0^l \dfrac{\sigma_3(y)}{E_3} dy = \dfrac{2(l - tanhkl)}{E_3 h_3} \tau_{20} = A_3 \tau_{20} \end{array}\right\} \tag{14}$$

where $A_1 = \dfrac{2(l - tanhkl)}{E_1 h_1}$, $A_2 = \dfrac{2(l - tanhkl)}{E_2 h_2}$, $A_3 = \dfrac{2(l - tanhkl)}{E_3 h_3}$.

After substituting Eqs. (14) in (13) we find:

$$\left.\begin{array}{l} \tau_{10} = \dfrac{A_2(\alpha_1 - \alpha_2)\Delta t l + A_2(\alpha_1 - \alpha_3)\Delta t l}{(A_1 + A_2)A_3 + A_1 A_2} \\[4mm] \tau_{20} = \dfrac{(A_1 + A_2)(\alpha_1 - \alpha_3)\Delta t l - A_1(\alpha_1 - \alpha_2)\Delta t l}{(A_1 + A_2)A_3 + A_1 A_2} \end{array}\right\} \tag{15}$$

With known $\tau_{10}$, $\tau_{20}$, by Eqs. (7) and Eqs. (12), it is possible to find the interface shear stresses and peel stresses.

## 3. Results and Discussion

### 3.1. Residual stresses in the ZrO$_2$/(ZrO$_2$+Ni) sandwich ceramics

The interface stresses computed by FEM and our model are compared as shown in Fig.4-5. We can see that the theoretical model results agree well with the results obtained by the commercial finite element package. The distribution of the shear stress is zero in the center and gradually reaches its peak near the edge, while the peel stress is compressive stress in the center and increasingly turned to tensile stress near the edge. The residual shear stress and peel stress in the interface are harmful to the shear strength and compressive strength of the ZrO$_2$/(ZrO$_2$+Ni) sandwich ceramics, respectively. These results agree well with the experimental results in [11], which conducted the shear test and compressive test of the $ZrO_2/(ZrO_2+Ni)$ sandwich ceramics and found both had a degradation compared with the porous $ZrO_2$ ceramic without interfaces.

Fig.6 shows the depth profile of the residual compressive stress in the middle layer (porous $ZrO_2$). We can see that the residual compressive stress runs through the middle $ZrO_2$ layer and the peak residual compressive stress (peak value reaches 88.12MPa for $h_2$=3mm) occurs near the interface, then the residual compressive stress decreases gradually as approaching the center. The residual compressive stress can make the crack deflect or bifurcate, preventing the further propagation of the crack, therefore, improving the bending strength of the $ZrO_2/(ZrO_2+Ni)$ sandwich ceramics. This is also in accordance with the experimental results in [11], where the bending test of the $ZrO_2/(ZrO_2+Ni)$ sandwich ceramics were performed and it was found that the bending strength had an increment up to 92% compared with the porous $ZrO_2$ ceramic which did not have the residual compressive stress.

### 3.2. Effects of the thickness ratio of the transitional layer to the middle layer

In order to study the effect of the thickness ratio of the transitional layer to the middle layer, which is defined as $p = h_2/h_3$, on the level of the residual stresses, the thickness of the middle porous $ZrO_2$ layer is kept to be constant ($h_3$=10mm), and different thicknesses of the transitional layer are investigated, including $h_2$=1mm, 2mm, 3mm and 4mm, with the mechanical loading layer being $h_1$=11mm, 10mm, 9mm and 8mm, respectively.

Fig.7 shows the computed middle layer residual compressive stress, interface residual peel stress, and interface residual shear stress. We can see that all stress components increase as the thickness of the transitional layer decrease. Considering that the interface residual peel stress, interface residual shear stress increase relatively higher while the middle layer residual compressive stress only has a slightly increase when the thickness ratio turns from 0.2 to 0.1, to suppress the interface residual peel stress, interface residual shear stress while keep a relatively bigger middle layer residual compressive stress, 0.2 is the most favorable thickness ratio of the transitional layer to the middle layer.

### 3.3. Effects of the number of transitional layers

To study the effect of the number of transitional layers on the level of residual stresses, the different number of transitional layers have been investigated, including one transitional layer (TZ3Y-15vol% Ni), two transitional layers (TZ3Y-10vol% Ni, TZ3Y-20vol% Ni) and three transitional layers (TZ3Y-7.5vol% Ni, TZ3Y-15vol% Ni, TZ3Y-22.5 vol% Ni). The total thickness and the porosity of transitional layers are kept constant, i.e. 2mm and 15%, respectively).

Fig.8 shows the computed middle layer residual compressive stress, interface residual peel stress, and interface residual shear stress. We can see that when the number of transitional layers turns form one layer to two layers, the middle layer residual compressive stress, interface residual peel stress both increase, while the interface residual shear stress gets quite small changes. When the number of transitional layers turns form two layers to three layers, interface residual shear stress decreases, but the middle layer residual compressive stress increases relatively small, moreover, the depth of the residual compressive stress of the middle layer decreases. Considering the complexity it brings when there are many layers to fabricate and process, five layers (one transitional layer) is the most favorable layer number.

## 4.    Conclusions

A simplified analytical model considering the overall deformation compatibility has been established to compute the interlaminar residual stresses, which agrees well with the results obtained by the commercial finite element package. The improvement of the flexural strength of the $ZrO_2/(ZrO_2+Ni)$ sandwich ceramics is due to the residual compressive stress produced in the middle $ZrO_2$ layer, which can make the crack deflect or bifurcate, preventing the further propagation of the crack. The degradation of the compressive strength and shear strength of the $ZrO_2/(ZrO_2+Ni)$ sandwich ceramics is mainly because of the interfacial residual stresses.

An optimal design has been done on the $ZrO_2/(ZrO_2+Ni)$ sandwich ceramics, the effects of the thickness ratio of the transitional layer to the middle layer, and the number of transitional layers on the mechanical properties are researched. An optimal structure of the $ZrO_2/(ZrO_2+Ni)$ sandwich ceramics is 0.2 for the thickness ratio of the transitional layer to the middle layer, and five layers in total layer numbers.

## Acknowledgments

The authors are grateful for the support by the National Natural Science Foundation of China (Grant Nos. 11102003, and 11472038).

**Figures and Tables**

Table1    Temperature-dependent material properties [14-16]

| Material | T(°C) | 25 | 300 | 600 | 800 | 1000 | 1127 |
|---|---|---|---|---|---|---|---|
| $ZrO_2$ | E(GPa) | 198.67 | 187.82 | 166.76 | 149.73 | 131.23 | 118.94 |
| $ZrO_2$ (30% porosity) | E(GPa) | 115.43 | 109.13 | 96.89 | 86.99 | 76.24 | 69.11 |
|  | $\alpha(10^{-6}/K)$ | 8 | 8.60 | 9.20 | 9.60 | 10 | 10.254 |
|  | $\mu$ | 0.315 | 0.315 | 0.315 | 0.315 | 0.315 | 0.315 |
| Ni | E(GPa) | 205 | 204.81 | 176.82 | 158.16 | 139.50 | 127.65 |
|  | $\alpha(10^{-6}/K)$ | 12.73 | 16.45 | 17.93 | 19.15 | 21.07 | 22.51 |
|  | $\mu$ | 0.3 | 0.3 | 0.3 | 0.3 | 0.3 | 0.3 |
| $ZrO_2$+15%Ni (15% porosity) | E(GPa) | 153.88 | 146.75 | 129.72. | 116.39 | 102.12 | 92.70 |
|  | $\alpha(10^{-6}/K)$ | 8.71 | 9.78 | 10.51 | 11.03 | 11.66 | 12.09 |
|  | $\mu$ | 0.313 | 0.313 | 0.313 | 0.313 | 0.313 | 0.313 |
| $ZrO_2$+30%Ni (12% porosity) | E(GPa) | 163.08 | 156.85 | 138.04 | 123.79 | 108.72 | 98.83 |
|  | $\alpha(10^{-6}/K)$ | 9.41 | 10.95 | 11.82 | 12.46 | 13.32 | 13.93 |
|  | $\mu$ | 0.311 | 0.311 | 0.311 | 0.311 | 0.311 | 0.311 |

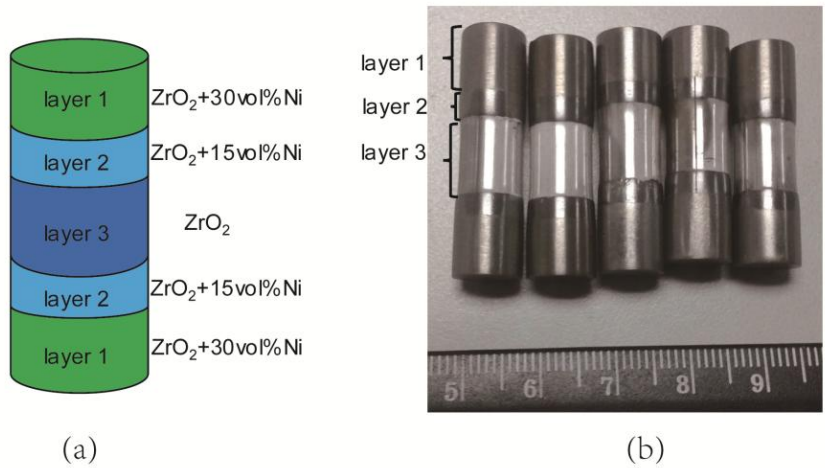

**Fig.1    Schematic composition and optical photograph of the fabricated ZrO₂/(ZrO₂+Ni) sandwich ceramics.**

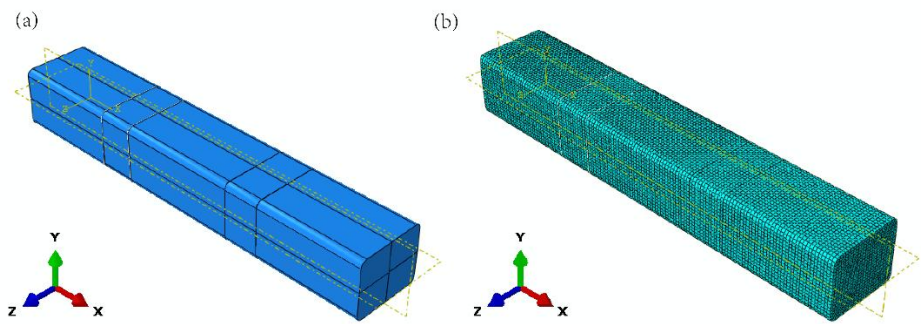

**Fig.2    Plot of the finite element model and the divided meshes of the ZrO₂/(ZrO₂+Ni) sandwich ceramics.**

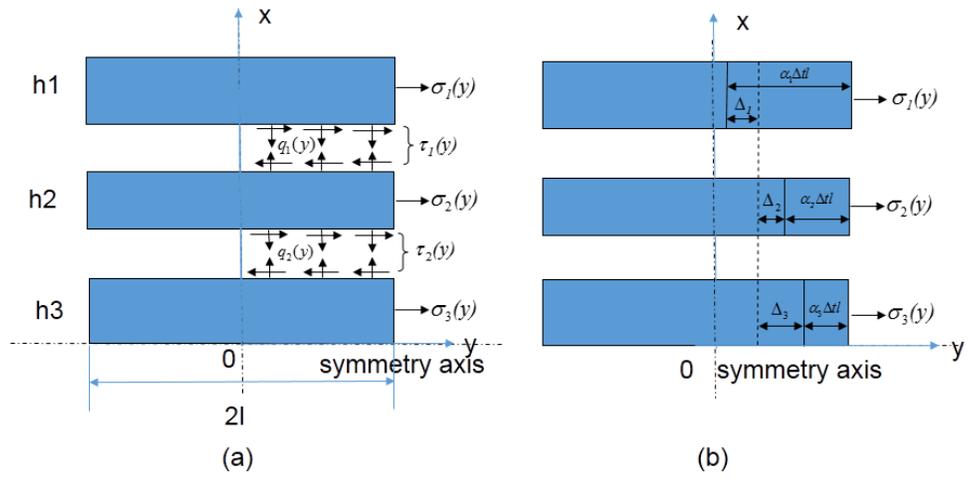

**Fig.3** **(a) The simplified stress analysis model of the ZrO₂/(ZrO₂+Ni) sandwich ceramics, (b) The displacement compatibility condition of the ZrO₂/(ZrO₂+Ni) sandwich ceramics.**

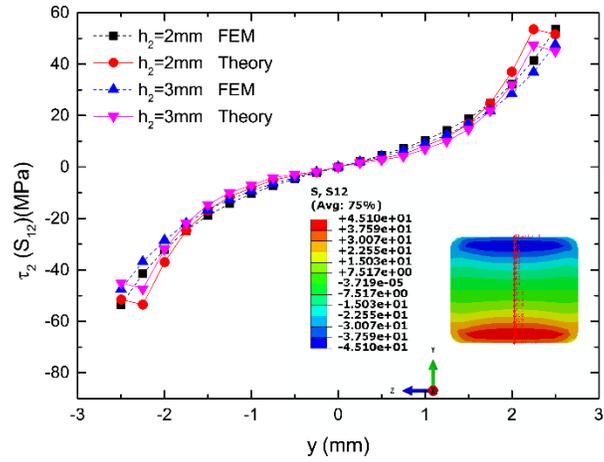

**Fig.4    The shear stress profile in the interface between the middle layer and the transition layer. Here, y refers to the path (red line on the FEM contour).**

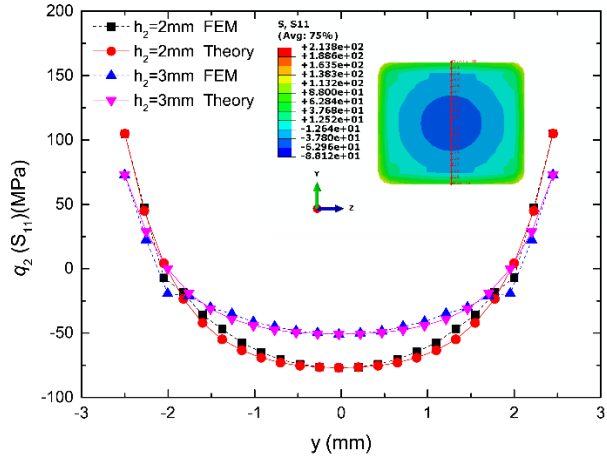

**Fig.5** **The peel stress profile in the interface between the middle layer and the transition layer. Here, y refers to the path (red line on the FEM contour).**

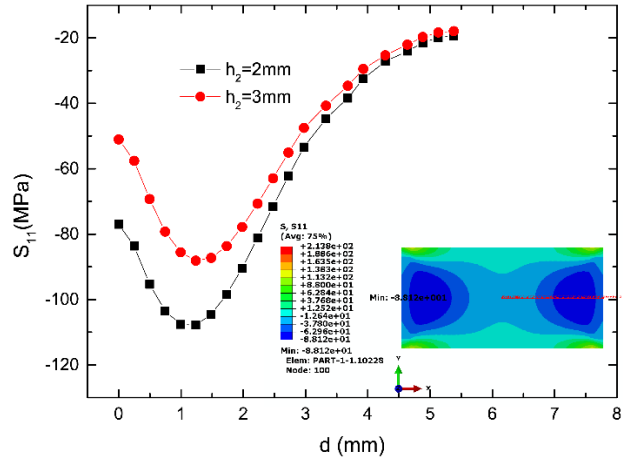

**Fig.6** **The depth profile of the residual compressive stress in the middle layer (ZrO₂). Here, d refers to the distance from the interface of the middle layer, also shown by the red line on the FEM contour.**

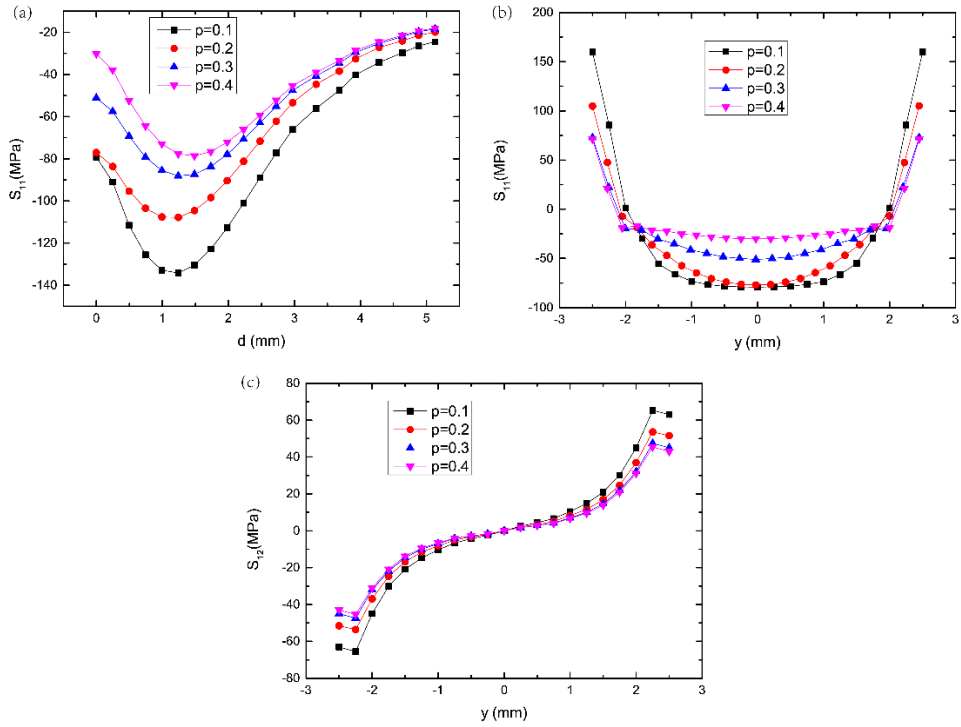

**Fig.7** **The relationship between the residual stresses and the thickness ratio of the transitional layer to the middle layer: (a) residual compressive stress in the middle layer, (b) peel stress in the interface of the middle layer, (c) shear stress in the interface of the middle layer.**

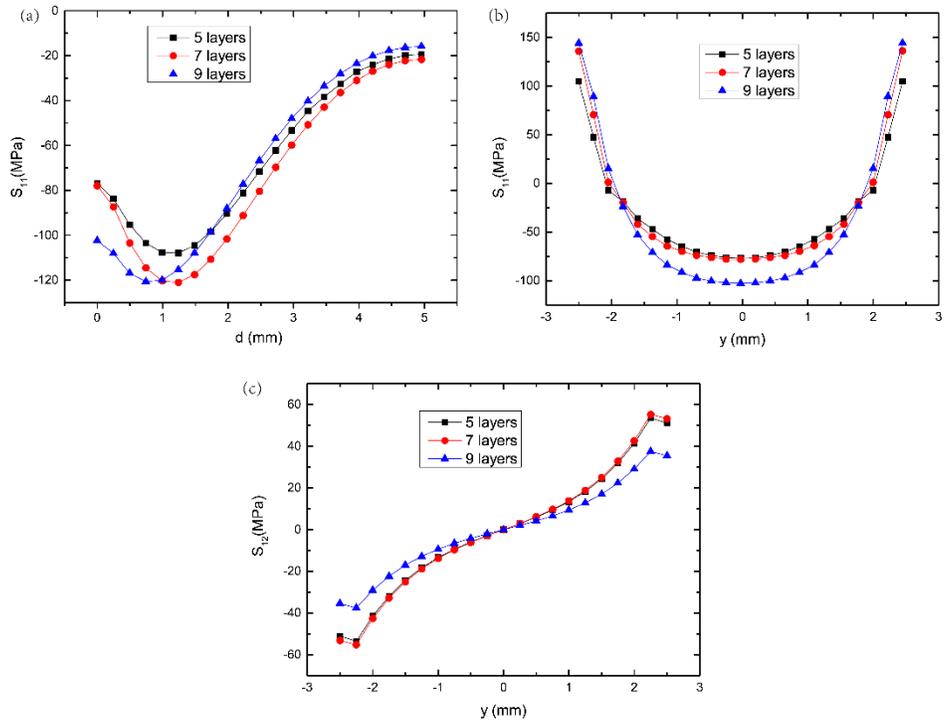

**Fig.8 The relationship between the residual stresses and the number of transition layers: (a) residual compressive stress in the middle layer, (b) peel stress in the interface of the middle layer, (c) shear stress in the interface of the middle layer.**